\documentclass[iop]{emulateapj}
\usepackage{newtxtext,newtxmath}
\usepackage[usenames]{color}
\usepackage{hyperref}

\hypersetup{colorlinks=true,linkcolor=black,citecolor=blue,filecolor=cyan,urlcolor=magenta}

\newcommand{\cM}{{\cal{M}}}

\newcommand{\OOc}{\Omega/\Omega_{\rm C}}

\newcommand{\mybibitem}[3]{\bibitem[{#1}({#2})]{#3}}
\newcommand{\mybibthree}[4]{\bibitem[{#2}({#3}){#1}]{#4}}
\submitted{Received 2018 July 11; revised 2018 August 6;
  accepted 2018 August 11; published 2018 August 24}
\journalinfo{Published in The Astrophysical Journal Letters, 864:L3, 2018 September
1}

\shorttitle{The Minimum Mass of Rotating MS Stars and its Impact on the Nature
  of eMSTOs} 
\shortauthors{Goudfrooij et al.}

\begin{document}

\title{The Minimum Mass of Rotating Main Sequence Stars and its Impact on the
  Nature of Extended Main Sequence Turnoffs in Intermediate-Age Star
  Clusters in the Magellanic Clouds\altaffilmark{1}}
\definecolor{MyBlue}{rgb}{0.3,0.3,1.0}
\author{Paul Goudfrooij$^2$, L\'eo Girardi$^3$, Andrea Bellini$^2$,
  Alessandro Bressan$^4$, Matteo Correnti$^2$, and Guglielmo Costa$^4$} 
\affil{$^2$ Space Telescope Science Institute, 3700 San Martin Drive,
  Baltimore, MD 21218, USA;
  \href{mailto:goudfroo@stsci.edu}{\color{MyBlue}goudfroo@stsci.edu},
       {\color{MyBlue}bellini@stsci.edu, correnti@stsci.edu}  \\ 
$^3$ Osservatorio Astronomico di Padova -- INAF, Vicolo dell'Osservatorio 5,
I-35122 Padova, Italy; {\color{MyBlue}leo.girardi@inaf.it} \\
$^4$ SISSA, via Bonomea 365, I-34136 Trieste, Italy;
{\color{MyBlue}sbressan@sissa.it, gcosta@sissa.it}}

\altaffiltext{1}{Based on observations with the NASA/ESA {\it Hubble
    Space Telescope}, obtained at the Space Telescope Science
  Institute, which is operated by the Association of Universities for
  Research in Astronomy, Inc., under NASA contract NAS5-26555} 


\begin{abstract}
Extended main sequence turn-offs (eMSTOs) are a common feature in 
color-magnitude diagrams (CMDs) of young and intermediate-age star clusters in
the Magellanic Clouds. The nature of eMSTOs is still debated. The most
popular scenarios are extended star formation and ranges
of stellar rotation rates. Here we study implications of a kink feature in the
main sequence (MS) of young star clusters in the Large Magellanic Cloud
(LMC). This kink shows up very clearly in new 
\emph{Hubble Space Telescope} observations of the 700-Myr-old cluster NGC~1831,
and is located below the region in the CMD where multiple or wide MSes,
which are known to occur in young clusters and thought to be due to
varying rotation rates, merge together into a single MS. The kink 
occurs at an initial stellar mass of $1.45 \pm 0.02\;M_{\odot}$; we posit
that it represents a lower limit to the mass below which the effects of
rotation on the energy output of stars are rendered negligible at the
metallicity of these clusters.    
Evaluating the positions of stars with this initial mass in CMDs of
massive LMC star clusters with ages of $\sim$\,1.7~Gyr that feature wide
eMSTOs, we find that such stars are located in a region where the eMSTO is
already significantly wider than the MS below it. This strongly suggests that
stellar rotation \emph{cannot} fully explain the wide extent of eMSTOs in
massive intermediate-age clusters in the Magellanic Clouds. 
A distribution of stellar ages still seems necessary to explain the
eMSTO phenomenon.
\end{abstract}

\keywords{stars: rotation --- globular clusters: individual (NGC 1783, NGC
  1806, NGC 1831, NGC 1846, NGC 1866) --- globular clusters: general ---
  Magellanic Clouds}   



\section{Introduction} 
\label{s:intro}

One of the significant discoveries enabled by the high-precision
photometry made possible with the Advanced Camera for Surveys (ACS) and the Wide
Field Camera 3 (WFC3) onboard the \emph{Hubble Space Telescope (HST)} was that
of extended main sequence turnoffs (hereafter eMSTOs) in massive
intermediate-age ($1-2$~Gyr old) star clusters in the Magellanic Clouds 
\citep{macbro07,milo+09,goud+09}. 
Such eMSTOs are much wider than the expectation of a simple stellar population
(SSP) in conjunction with photometric uncertainties and stellar binarity. 
Furthermore, several eMSTO clusters feature a faint extension to the
red clump of core helium burning giants, indicating the presence of a
significant range of stellar core masses at the He flash
\citep{gira+09,rube+11}.  

The nature of the eMSTO phenomenon is still debated. The perhaps simplest 
explanation is an age spread of up to several $10^8$ yr within these clusters
\citep[e.g.,][]{mack+08a,milo+09,goud+14,goud+15}.  
In this ``age spread'' scenario, the shape of the star density distribution
across the eMSTO is thought to reflect the combined effects of 
the histories of star formation and cluster mass loss due to strong cluster
expansion following the death of massive stars in the central regions
\citep[][hereafter G+14]{goud+14}.
Support in favor of this scenario was provided by G+14 by means of a correlation
between MSTO width and central escape velocity, which is a proxy for the
cluster's ability to retain and/or accrete gas during its early evolution. G+14
also reported a strong correlation between the fractional numbers of stars in
the bluest region of the MSTO and those in the faint extension of the red clump,
as expected from an age spread. 

The leading alternative explanation of the eMSTO phenomenon is that it is 
caused by a spread of stellar rotation rates. Rotation lowers the luminosity and
effective temperature at the stellar surface and \citet{basdem09} argued that
this, when combined with projection effects, could cause eMSTOs similar to those
observed. More recent studies revealed an opposite effect of stellar rotation,
namely a longer main sequence (MS) lifetime due to internal mixing
\citep{gira+11,geor+14}, thus mimicking a \emph{younger} age. 
Significant support for the rotation scenario was provided by \emph{HST}
observations of younger clusters in the Large Magellanic Cloud (LMC) with ages
of $\sim100-300$ Myr 
\citep[e.g.,][]{corr+15,corr+17,milo+15,milo+17}.
These revealed the presence of eMSTOs combined with broadened or split MSes which are
predicted to occur at these ages when a significant fraction of stars has rotation
rates $\Omega$ in excess of $\sim80$\% of the critical rate ($\Omega_{\rm C}$)
according to the Geneva {\sc syclist} isochrone models of \citet{geor+14},
whereas it cannot easily be explained by age spreads
\citep[see][]{dant+15,milo+16,corr+17}.   
Finally, several stars in the MSTOs of young massive LMC clusters are now known
to be strong H$\alpha$ emitters and thought to constitute equator-on Be stars
that are rapidly rotating at $\OOc\ga0.5$ \citep{bast+17,corr+17,dupr+17,milo+18}. 

These recent findings suggest that stellar rotation is part of the
explanation of the eMSTO phenomenon \citep[see also][]{mart+18,mari+18}. 
However, there are relevant indications that other effects are at play as well. 
For example, the distributions of stars across the eMSTOs of several young massive
clusters are \emph{not} consistent with a coeval population of stars
encompassing a range of rotation rates (as represented by the {\sc syclist}
model predictions). Specifically, the number of stars on the ``red'' side of the
MSTO is significantly higher than that expected if the ``blue'' side of the MSTO  
constitutes the bulk of the rotating stars in a coeval population of stars,
hinting at the presence of an age spread in addition to rotation
\citep[see][]{corr+17,milo+16,milo+17,milo+18}. 
Furthermore, \citet{goud+17} showed that the
distribution of stars across the eMSTOs of two massive LMC clusters with an age
of $\sim1$~Gyr cannot be explained 
solely by a distribution of rotation rates according to
the {\sc syclist} models, unless the orientations of rapidly rotating
stars are heavily biased toward an equator-on configuration \emph{in both
  clusters}, which is statistically highly improbable.

Finally, the {\sc syclist} isochrones only cover stellar masses
$\cM\geq1.7\;M_{\odot}$, because this is the mass below which magnetic braking
occurs due to the presence of a convective envelope, complicating the
calculations of the influence of rotation to stellar evolutionary models.
Notwithstanding this complexity, the magnetic braking for stars with
$\cM/M_{\odot}<1.7$ is expected to decrease the effects of rotation, which
should therefore result in MSTOs that get narrower with increasing age for
intermediate-age LMC clusters with ages $\ga1.2$~Gyr. However, this is
inconsistent with observations in that the LMC clusters with the widest known
MSTO's have ages in the range $1.6-1.8$~Gyr (see, e.g., G+14). These results
cast doubt on the assertion that stellar rotation is the only cause of eMSTOs
in intermediate-age clusters.   

In the context of this debate, one may wonder if there might be
model-independent features in the CMD that could yield a direct indication as to
whether stellar rotation can fully explain the morphology of eMSTOs in massive
intermediate-age clusters. 
In this paper, we study the implications of one such feature, namely an obvious
kink in the MS of clusters younger than $\approx800$ Myr, which provides an
empirical measurement of the stellar mass below which the influence of
rotation to the width of the MS is negligible (for the metallicity of these
clusters).

\section{Observations and Data Reduction}
\label{s:obs}

Our attention to this kink in the MS of young LMC clusters was drawn by our
recent \emph{HST} observations of NGC~1831, which was the target of \emph{HST}
observing program GO-14688 (PI:~P.~Goudfrooij), using the UVIS channel of the Wide
Field Camera 3 (hereafter WFC3/UVIS). Multiple dithered images were taken
through the F336W and F814W filters, with total exposure times of 4180 s and
1480 s, respectively. 
Data analysis was carried out on the flat-fielded {\tt *\_flc.fits} images that
were corrected for charge transfer inefficiency. Stellar photometry measurements
were done with point spread function (PSF) fitting, using the ``effective PSF'' 
(ePSF) package for WFC3/UVIS (J.~Anderson, private communication), which is
based on the ePSF package for the ACS/WFC camera described in
\citet{andkin06}. More details on this dataset and the photometry will be
provided in a separate paper (M.~Correnti et al., in preparation), 
but all data used in this paper can be obtained from MAST at 
  \dataset[10.17909/T9P41K]{https://doi.org/10.17909/T9P41K}.  

In order to minimize and illustrate the contamination by LMC field stars, we
extract a circular ``cluster region'' containing the stars within a projected
effective radius of the cluster center ($r_{e}=8.2$ arcsec,
\citealt{mclvdm05}). For reference, we also extract a ``background region'' with
the same area as the cluster region, but located near the edge of the image
furthest away from the cluster center.

\begin{figure*}
  \centerline{
    \includegraphics[width=1.0\textwidth]{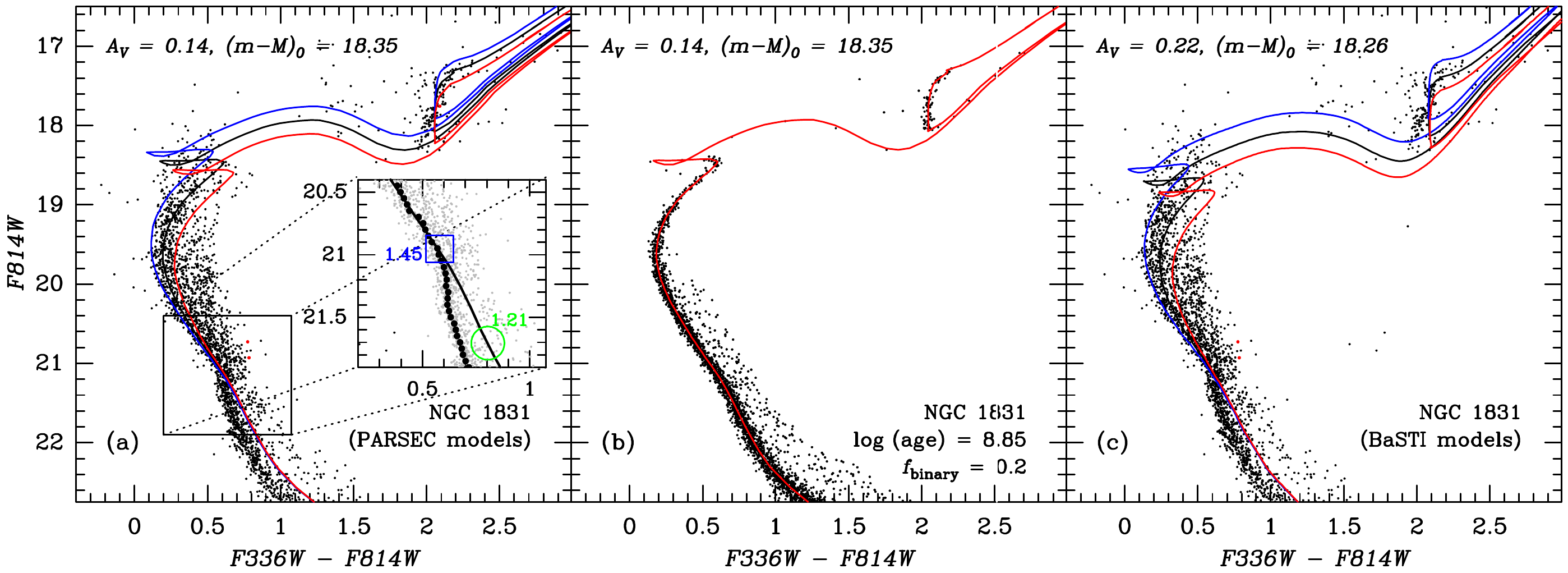}
  }
\caption{\emph{Panel (a)}: {\it F814W} vs.~${\it F336W}-{\it F814W}$ CMD of
  NGC~1831 within its effective radius, along with PARSEC isochrones for $Z=0.008$ and
  log~(age/yr)~=~8.80, 8.85, and 8.90 (in blue, black, and red,
  respectively). Values for $A_V$ and $(m-M)_0$ are indicated in the legend. 
  The (very few) red dots indicate stars in the ``background region'' discussed
  in the text.
  Note the obvious ``MS kink'' at ${\it F814W}\sim21$.
    The inset shows the kink region in which the
  individual stars are shown in grey, the black line is the PARSEC isochrone for
  log~(age/yr)~=~8.85, and the black circles represent the MS fiducial as a
  function of {\it F814W} magnitude. 
  The positions of stars with
  $\cM/M_{\odot}=1.45$ and $1.21$ are highlighted using blue and green markers, 
  respectively.  
  \emph{Panel (b)}: Synthetic PARSEC cluster with $Z=0.008$,
  log~(age/yr)~=~8.85, binary fraction of 0.2, and photometric uncertainties
  taken from the observations. 
  \emph{Panel (c)}: Similar to panel (a), but now showing 
  BaSTI isochrones for $Z=0.0057$ and ages of 0.6, 0.7, and 0.8 Gyr
  (in blue, black, and red, respectively). 
  } 
\label{f:CMD1831}
\end{figure*}

The CMD of NGC 1831 and the background region are shown in panel (a) of
Figure~\ref{f:CMD1831}. Note that the contamination by field stars is
negligible.
For comparison purposes, panel (b) of Figure~\ref{f:CMD1831} shows a CMD of a
synthetic cluster based on a (non-rotating) PARSEC v1.2 isochrone
\citep{bres+12} with $Z=0.008$ and log~(age/yr)~=~8.85. The simulation used a
\citet{salp55} initial mass function, a binary fraction of 0.2,
and photometric uncertainties taken from the observations.  
Note that the MSTO of NGC 1831 is significantly more extended than the
expectation of a SSP in conjunction with photometric uncertainties, as found for 
several other young and intermediate-age clusters in the LMC. While the
morphology of the eMSTO and upper MS of NGC~1831 will be studied in detail in
the context of the effects of 
stellar rotation in a separate paper (M.~Correnti et al., in preparation), we
focus here on the obvious kink in the MS of NGC~1831 at ${\it F814W}\sim21$. 
Above this kink, the MS broadens into a fan-like shape
which is likely due (at least in part) to a range of stellar
rotation rates and rotation axis orientations in the cluster. Conversely, below
this kink, the single-star MS emerges as narrow as would be expected for a SSP,
while exhibiting a sharp curve downward in the {\it F814W} vs.~${\it F336W}-{\it
  F814W}$ CMD which is not represented well by the isochrone model.

\section{The ``Kink'' in the MS of Young Clusters} 
\label{s:MSkink}

\begin{figure*}
  \centerline{
    \includegraphics[width=15.35cm]{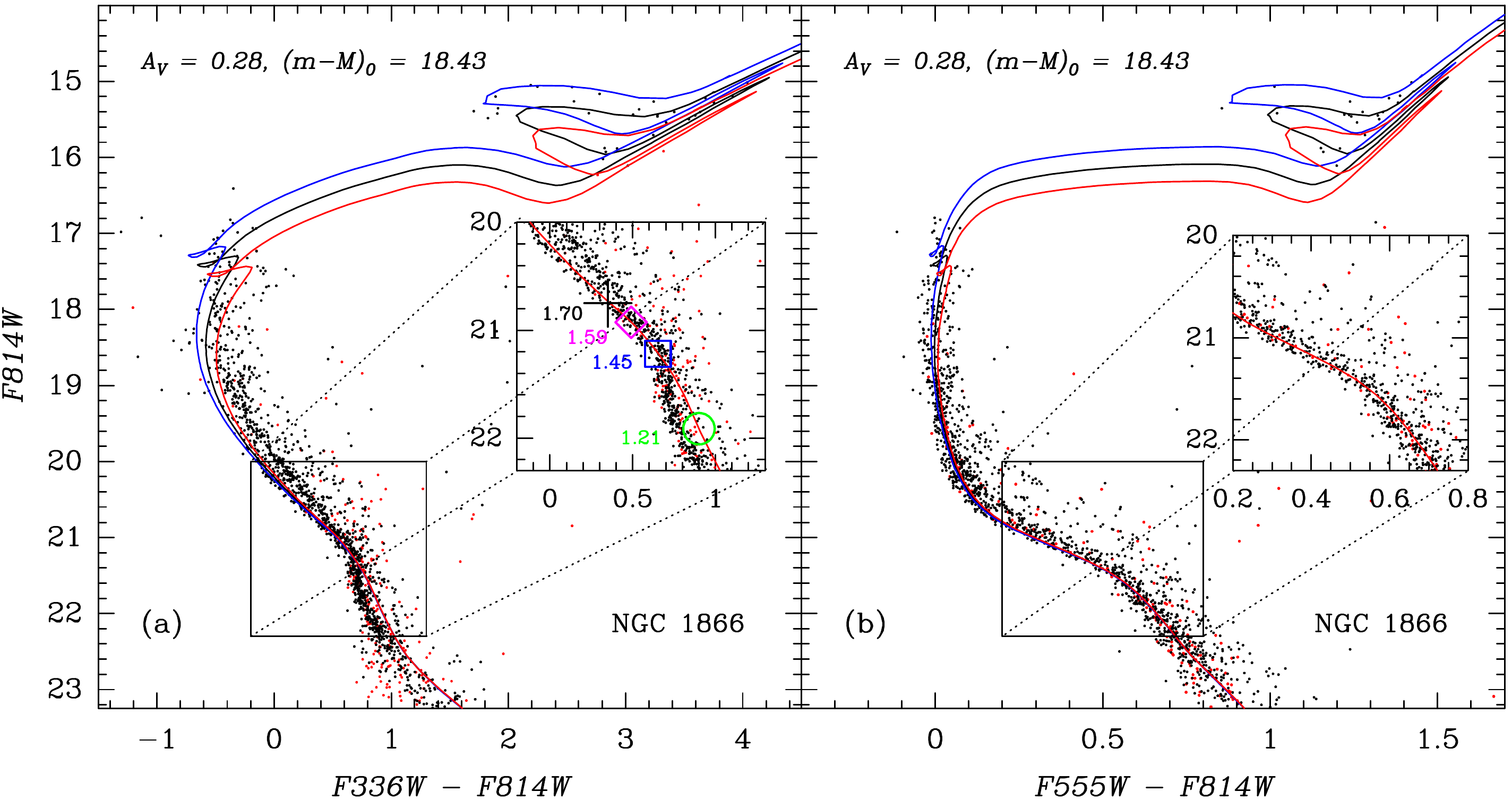}
  }
\caption{\emph{Panel (a)}: Similar to panel (a) of Figure~\ref{f:CMD1831}, but
  now for NGC~1866.  
  The PARSEC isochrones are for for $Z=0.008$ and log~(age/yr)~=~8.24, 8.30,
  and 8.36 in this case. 
   The inset highlights the
  positions of stars with $\cM/M_{\odot}=1.70$, 1.59, 1.45, and 1.21 using
  labels and markers in black (plus), magenta (diamond), blue
  (square), and green (circle), respectively. 
  \emph{Panel (b)}: Similar to panel (a), but now using the
  ${\it F555W}-{\it F814W}$ color. Note that the MS below the kink at
  ${\it F814W}\sim21.2$ is much better fit by the isochrones in
  ${\it F555W}-{\it F814W}$ than in ${\it F336W}-{\it F814W}$. 
  See discussion in Section~\ref{s:MSkink}.
  }
\label{f:CMD1866}
\end{figure*}

This MS kink is a general feature among young clusters in the LMC: 
\citet{milo+18} studied {\it F814W} vs.~${\it F336W}-{\it F814W}$ CMDs of
13 LMC clusters with ages between $\sim40$~Myr and 1.0~Gyr, and
found that all of them feature an eMSTO along with an MS that is wide and/or
split above a kink-like feature similar to that seen in NGC~1831. The narrow
single-star MS emerging below this kink can also be seen in the CMDs by
\citet{milo+18}, albeit typically at lower signal-to-noise ratio than in the CMD
of NGC~1831 shown here. The exception to the latter is the case of NGC~1866, for
which deep images were taken in programs GO-14204 (PI: A.~P.~Milone) and 
GO-14069 (PI: N.~Bastian), which we downloaded from the \emph{HST} archive and
processed in the same way as that described above for NGC~1831. CMDs for
NGC~1866 are shown in Figure~\ref{f:CMD1866}.  

To determine stellar parameters at the location of the MS kink, we compare the
data of NGC~1831 and NGC~1866 with non-rotating PARSEC v1.2S isochrones. We
fit the latter to the {\it F814W} vs.~${\it F336W}-{\it F814W}$ CMDs above the
MS kink in a way similar to \citet{milo+18},    
i.e., interpreting the blue edge of the MS above the kink as the MS of non-rotating
stars in these clusters, but now also taking the positions of the
core-helium-burning stars into account. 
It can be seen in panel (a) of Figures~\ref{f:CMD1831} and \ref{f:CMD1866} that
this way of fitting isochrones to these CMDs results in non-optimal 
fits to the single-star MS below the kink in the sense that the isochrones indicate
${\it F336W}-{\it F814W}$ colors redder than the data. However, the {\it F814W}
vs.~${\it F555W}-{\it F814W}$ CMD is actually fit very well by the same
isochrones (above \emph{and} below the MS kink), as can be seen for NGC~1866 in
panel (b) of Figure~\ref{f:CMD1866}.
It therefore seems that the isochrones may have trouble describing the
effective temperature of MS stars below this kink, possibly in
conjunction with problems with the synthetic spectra used to convert the
isochrones into colors involving passbands at shorter wavelengths, with
adequate precision. 
  This problem, which is discussed further below, is not
  restricted to the PARSEC isochrone models which we use here as an
  example;  several other popular models share the same issue. This is
  illustrated in panel (c) of Figure~\ref{f:CMD1831} for BaSTI \citep{piet+04}
  isochrones for which magnitudes were transformed to the \emph{HST/WFC3} passbands 
  following \citet{goud+17}.\footnote{We also verified this for 
  the Geneva models \citep{mowl+12}  as well as those of Y$^2$
  \citep{yi+01,dema+04}, Dartmouth \citep{dott+08} and MIST \citep{choi+16}.}

To determine the location of the kink, we construct MS fiducials that describe
the peak of the ${\it F336W}-{\it F814W}$ color distribution
in the range $20.5\leq{\it F814W}\leq22.0$, using
magnitude bins of 0.10 mag. The peak colors are taken to be the maximum of their
kernel density distributions, using an Epanechnikov kernel with adaptively chosen
bandwidth \citep{silv86}. The kink brightness is then defined as
the {\it F814W} magnitude beyond which the position of the MS fiducial in the CMD
is systematically below that of the best-fitting isochrone.
This is illustrated for NGC~1831 in the inset in Figure~\ref{f:CMD1831}a where
the kink is identified at ${\it F814W}=20.95\pm0.03$.
Using this method, and assuming the values of $m-M$ and $A_V$ shown in
  Figures~\ref{f:CMD1831} and \ref{f:CMD1866}, we find that the kink
  occurs at $M_{\it F814W}^{0}=2.46\pm0.03$ and $2.62\pm0.03$ for
  NGC~1831 and NGC~1866, respectively. Using linear interpolation
within the isochrones, these luminosities correspond to initial stellar masses
$\cM/M_{\odot}=1.45\pm0.02$ and $1.45\pm0.02$, respectively. 
This mass is in the range where stellar structure undergoes significant
changes between radiative and convective modes. Specifically, core convection is
thought to occur for stars with $\cM/M_{\odot} \ga 1.3$
\citep[e.g.,][]{egge+08}, while stars with $\cM/M_{\odot}\la1.7$ have 
convective envelopes, featuring magnetized winds which shed angular momentum
that the star may have built up during its formation era \citep[e.g.,][]{geor+14}.

The shape and location of this MS kink, in conjunction with the fact that it is
not well described by isochrone models, strongly suggests that its nature
is related to the sudden onset of strong convection in the outer layers of
stars with $\cM\simeq1.45\;M_{\odot}$. In most isochrone models, the energy
transport in the convection zone in stellar envelopes is modeled using the
mixing-length theory (MLT) of \citet{bohm58}, along with some
degree of convective ``overshoot'' into the radiative region. 
In this theory, the mixing length $\alpha$ is calibrated to fit the solar radius
at the age of the Sun, which yields $\alpha$ values of order $1.5-2.0$ 
pressure scale heights. However, several pieces of evidence suggest that the
efficiency of mixing varies significantly as a function of
radius within stars. For example, the shape of Balmer line profiles of
the Sun and lower-mass stars requires much smaller mixing lengths in
the outer layers of stars ($\alpha\sim0.5$ rather than $1.5-2.0$; see
\citealt{fuhr+93}). It was 
pointed out by \citet{bern98} that this discrepancy can be resolved by applying
the full spectrum of turbulence (FST) convection model by \citet{canmaz91}. 
\citet{dant+02} compared model isochrones built with MLT and FST convection
models and found that the FST model yields a sudden change in the MS slope at a
location in the CMD that is very similar to that of the MS kink in NGC~1831,
whereas the MLT model does not show this kink (see their Figure~2). As shown by
\citet{dant+02}, this kink occurs at a stellar mass for which the convective
envelope suddenly reaches much deeper into the interior than it does at a mass
only 0.01~$M_{\odot}$ larger, thus causing a significant decrease of the
temperature dependence of stellar mass, $dT_{\rm eff}/dM$, with decreasing
stellar mass. 
  Since the $T_{\rm eff}$ at the MS kink is $\sim7800$ K, for which the stellar
continuum peaks at $\sim3700$ \AA\ according to Wien's law, this sudden
decrease of $dT_{\rm eff}/dM$ is measured most precisely using filters around
that wavelength with a wide baseline, such as 
${\it F336W}-{\it F814W}$; conversely, red colors like ${\it F555W}-{\it F814W}$
mainly measure the change in slope of the Rayleigh-Jeans tail of the spectrum,
providing a less precise measure of $T_{\rm eff}$. 

We posit that the MS kink seen in NGC~1831 and the younger LMC clusters studied
by \citet{milo+18} and references therein is associated with this sudden change
in the extent of the convective envelope at the metallicity of the young and
intermediate-age LMC clusters. Furthermore, since the width of the single-star
MS emerging down from the MS kink in NGC~1831 is fully consistent with a SSP, we
posit that the kink also represents an empirical measure of the stellar mass
below which rotation has no appreciable influence on the energy output of the
star. This is consistent with the sudden increase of the depth of the convective
envelope, given that the latter is believed to be the cause of magnetic braking
of angular momentum in stars.
Note that the stellar mass associated with this MS kink
is formally a \emph{lower limit} to the mass
associated with the effective onset of stellar rotation, since the level of
envelope convection needed to stifle the effects of rotation is not
known theoretically. In this context, we note that the single-star MS of
NGC~1866 is found to be narrowest at {\it F814W} = $20.95\pm0.05$,
corresponding to $\cM=1.59\pm0.03\;M_{\odot}$ (see inset in
Figure~\ref{f:CMD1866}a), which may turn out to be closer to the stellar
mass associated with the effective onset of stellar rotation than the
1.45~$M_{\odot}$ mentioned above.

\section{Comparison with Intermediate-age eMSTO Clusters} 
\label{s:compare}

\begin{figure*}
\centerline{\includegraphics[width=17.cm]{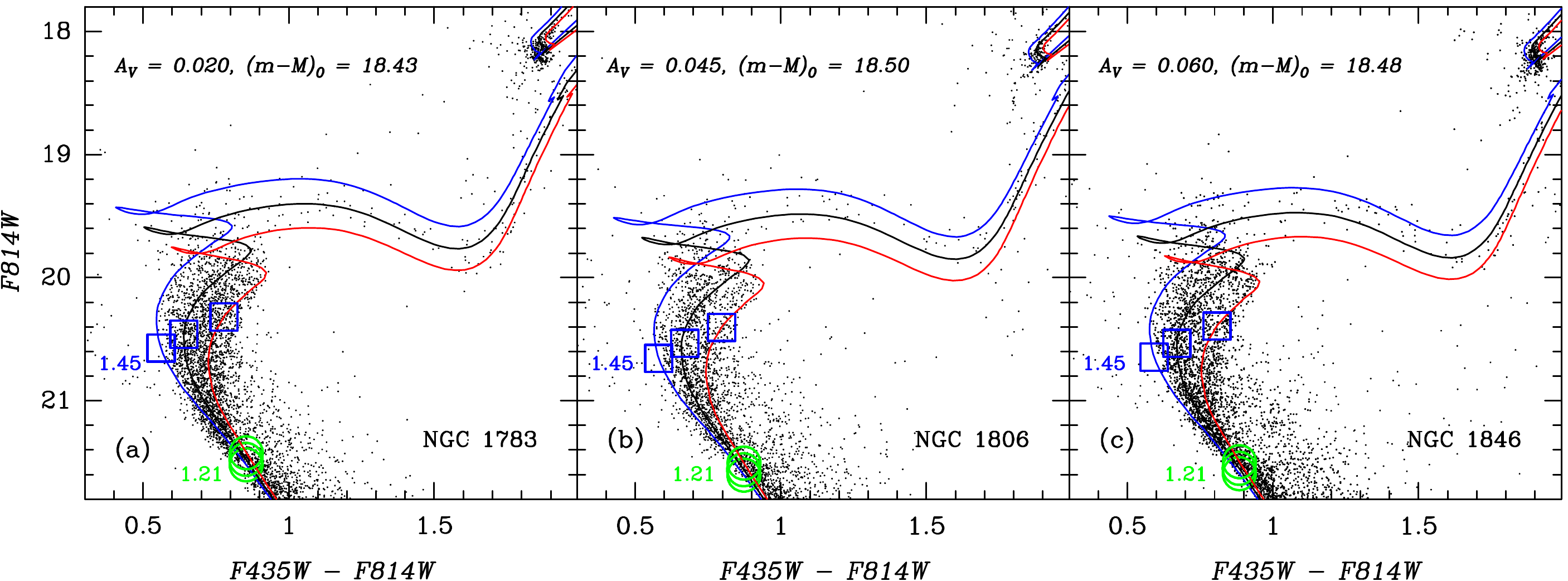}}
\caption{\emph{Panel (a)}: CMD of NGC~1783 within its core radius,
  taken from \citet{goud+11b}, along with PARSEC isochrones for $Z=0.008$ and
  log~(age/yr)~=~9.18, 9.24, and 9.30 (in blue, black, and red, 
  respectively). Values for $A_V$ and $(m-M)_0$ are indicated in the legend.
  The open blue squares and open green circles indicate the positions of stars with
  $\cM/M_{\odot}=1.45$ and 1.21, respectively, for each isochrone. 
  \emph{Panel (b)}:  similar to panel (a), but now for NGC~1806. 
  \emph{Panel (c)}:  similar to panel (a), but now for NGC~1846.
  See discussion in Section~\ref{s:compare}. 
  } 
\label{f:IntAgeCMDs}
\end{figure*}

To illustrate the impact of the analysis in the previous section to the
nature of eMSTO's in intermediate-age clusters, we compare the high-quality
\emph{HST/ACS} photometry of NGC~1783, NGC~1806 
and NGC~1846, three massive eMSTO clusters in the LMC with ages of $\sim1.7$~Gyr
\citep{mack+08a,milo+09,goud+11b}, with the PARSEC isochrones (i.e., 
the same models as those used to determine the stellar mass at the MS kink in
the previous section). These clusters were selected for this 
purpose because they feature the widest known eMSTOs among intermediate-age clusters
in the LMC.\footnote{The widest known eMSTO is that of NGC~419 in the SMC
  (e.g.,~G+14), but we do not select that cluster for this
  comparison due to uncertainties associated with its significantly lower metallicity.} 
The \emph{HST/ACS} photometry and the isochrone fitting procedure for these
three clusters was described before by \citet{goud+09,goud+11b}. 
Figure~\ref{f:IntAgeCMDs} shows the {\it F814W} vs.~${\it F435W}-{\it F814W}$
CMDs of these clusters along with three PARSEC isochrones for $Z=0.008$ and 
log~(age/yr)~=~9.18, 9.24, and 9.30, fitting the approximate blue end, center,
and red end of the eMSTOs of these clusters, respectively.   
The locations of stars with the same initial mass as that associated with the MS
kink in the younger LMC clusters, i.e., $\cM=1.45\;M_{\odot}$, are shown as
open squares on top of these isochrones. Note that \emph{the widening of the
  eMSTOs in these clusters starts at significantly lower stellar masses than that 
  associated with the MS kink in younger LMC clusters}.
Specifically, the width of the single-star MS becomes consistent with a SSP at
${\it F814W}\sim21.5$ for these three intermediate-age
clusters. This corresponds to an initial stellar mass of
$\cM=1.21\pm0.02\;M_{\odot}$, where the uncertainty represents the
dispersion of ages required to fit the distributions of stars across the MSTOs
of these clusters as well as the different values of $(m-M)_0$ and $A_V$ of the
three clusters. This initial mass is smaller by $\sim10\,\sigma$ than the
initial mass associated with the MS kink in LMC clusters with ages $\la700$
Myr, i.e., the location where their wide MS merges into a narrow single-star MS
whose width is consistent with a SSP. This strongly suggests that stellar
rotation is \emph{not} the only cause of the extended MSTOs in the massive
intermediate-age clusters. 

\section{Conclusion}
\label{s:conc}

While several recent studies have provided important evidence in
support of the stellar rotation scenario for the nature of eMSTOs in young
clusters in the LMC, our results show that the very wide MSTOs in the most
massive intermediate-age clusters with ages of $1.6-1.8$~Gyr cannot fully be
explained by rotation. This is simply because the MSTO in the latter clusters
widens at a stellar mass well below that associated with the MS kink in younger
clusters below which, as we showed above, the width of the MS of single stars is
significantly narrower than that above the MS kink, and consistent with that of
an SSP of non-rotating stars. These observations could be reconciled with the
rotation-spread hypothesis only if there is a mass range in between 1.45 and
1.21~$M_{\odot}$ for which the effects of rotation take longer than 700 Myr
(i.e., the age of NGC~1831) to produce a detectable effect at the stellar
surface. Although the mixing induced by rotation could in principle produce 
effects that increase with age, it appears unlikely that they would 
manifest themselves so clearly at an age of $\sim1.7$~Gyr, also since magnetic
braking is thought to be powerful in stars with strong envelope
convection (i.e., with $1.20\la\cM/M_{\odot}\la1.45$; see, 
  e.g., \citealt{vans+16}). 

This result constitutes new support for the age spread scenario on the nature
of eMSTOs of massive intermediate-age clusters. Nevertheless, the recent studies
of young clusters with eMSTOs have also clarified that rotation is part
of the solution as well, so that the age spreads derived by G+14 are rendered
overestimates \citep[see discussion in][]{goud+17}. Studies of   
the effects of rotation at $\cM<1.7\;M_{\odot}$ are currently
being pursued by different isochrone modelers including PARSEC and MIST,
which should yield relevant new insights in this context. 

\acknowledgments
We thank the referee for a constructive and helpful report which
benefited the paper. Support for this project was provided by NASA through grants
HST-GO-14688 and HST-AR-15023 from the Space Telescope Science Institute, which is
operated by the Association of Universities for Research in Astronomy, Inc.,
under NASA contract NAS5-26555.

%

\vspace{5mm}
\facility{\emph{Facilities:} HST(ACS, WFC3)}

\end{document}